# *Ab-initio and Critical behaviors of the perovskite CaMnO3 for solar cell applications*

H. Mahrouch [1], A. Jabar [2], S. Idrissi [1] and L. Bahmad [1]

[1] Laboratory of Conception and Systems (LCS), Faculty of Sciences, Mohammed V University in Rabat, Av. Ibn Batouta, B. P. 1014 Rabat, Morocco

[2] LPMAT, Faculty of Sciences Aïn Chock, Hassan II University of Casablanca, B.P. 5366 Casablanca, Morocco

Corresponding author: l.bahmad@um5r.ac.ma

## Abstract

In this work, we used the density functional calculation (DFT) implemented in the Quantum Espresso software, using the approximations (GGA, GGA+U) to illustrate the electronic and magnetic properties of the perovskite CaMnO3. It has been found that the CaMnO3 perovskite is stable in the G-AFM phase. When expecting the total and partial DOSs, a strong contribution of the d-Mn states has been outlined. The Coulomb correction U and the site exchange interaction J have been implemented and then we compared the two approximations GGA and GGA+U. It is found that the GGA+U method leads to more accurate results since this correction takes into account the bonding effects between different atoms. To complete this study we performed the simulations under Monte Carlo code based on the Metropolis algorithm. In fact, we have simulated the physical quantities: magnetization, susceptibility, and specific heat of the studied CaMnO3 material as a function of temperature.

1. Introduction

Materials with a perovskite structure have attracted significant interest for more than a decade due to their exceptional electrical, magnetic, and optical properties. These properties often exhibit sensitivity to external factors such as temperature, pressure, and phase changes, making them versatile and adaptable to a variety of applications. The perovskites were mainly synthesized and studied by Russian and Japanese researchers, they have a unique atomic structure typically ABO3, first described in 1830, named after the Russian mineralogist Lev Perovski. The compounds ABO3 form a family of fascinating compounds, where A is an alkali or alkaline earth metal, and B is a transition metal, which is currently forming a promising new class of materials (both elements A and B are cations, and O is an anion). These perovskite-like oxides have attracted much attention because they exhibit a range of interesting properties such as superconductivity, ferroelectricity, semi-conductivity, ionic conductivity, piezoelectricity, thermoelectricity, ferromagnetism, semi-metallic transport, and colossal magnetoresistance. [1, 2].

The material CaMnO3 is a fascinating example of the rich properties of perovskite materials and continues to attract the attention of researchers in materials science and condensed matter physics. The CaMnO3 is a complex compound with attractive magnetic and electrical properties. It is able to change its behavior depending on the changes in temperature and magnetic fields. This makes it a material of interest in science research and opens up promising prospects for industrial applications technology in the future. In many theoretical works, CaMnO3 is compared with SrMnO3, which has a 4H hexagonal structure at low temperatures and a cubic structure at high-temperature values [3-4]. All works studied the cubic perovskite structure. The study by Søndena et al. [4] for studying the electronic structure of the series of AMnO3 (A=Ca, Sr, Ba) using the GGA approximation in the cubic, 2H, and 4H-hexagonal structures for all compounds proved that CaMnO3 adopts the ideal cubic perovskite and was found to be G-AFM. Another theoretical work by Poeppelmeier et al. [5] mentioned that the CaMnO3 compound (t=0.987) adopts a distorted cubic perovskite structure (all apex sharing), and Søndena et al. [4] confirmed in their study that the tolerance factor predicts the cubic structure for CaMnO3 (t=0.94). Density Functional Theory (DFT) is a highly valuable and widely used computational method for accurately predicting the electronic and magnetic properties of a large number of materials. In our paper, we discuss the structural, magnetic, and electronic properties of CaMnO3 in the ideal cubic perovskite phase, by using the approximation GGA, and by the introduction of the U-Hubbard term in our calculations where we used GGA+U. In the first step, we examined the structural properties using the approximation GGA. Then, we perform a comparative study between GGA and GGA+U. In particular, we focus our research on the electronic and magnetic properties to find out more accurate results when using the correction U-Hubbard. We studied the ideal cubic perovskite structure in the ferromagnetic (Ferro) configuration, designated by the space group Pm-3m, and the G-type anti-ferromagnetic configuration (G-AFM), designated by the space group Fm-3m. We show the band structure and the density of states (DOS). Finally, we conclude this work with our conclusions.

On the other hand, Ref. [6] presents the thermoelectric properties of the perovskite $CsSnBr_3$ using two altered techniques. It is worth noting that the type of cations has a very slight impact on the optical characteristics of such materials [7]. Moreover, the electronic, optical, and thermoelectric properties of the $CsMF_3$ (M= Si or Ge) perovskites and the magnetic and magneto-optical properties of doped and co-doped CdTe with (Mn, Fe) have been illustrated in the references [8] and [9], respectively. In addition, the DFT and TDDFT methods have been performed to clarify the structural, electronic, and optical properties of the inorganic solar perovskites $XPbBr_3$ (X = Li or Na) [10]. In addition, the Monte Carlo method has been directed to highlight the critical behavior of several double perovskites such as $Sr_2VMoO_6$ [11], $Sr_2CrIrO_6$ [12], $Ba_2NiUO_6$ [13], $Sr_2RuHoO_6$ [14], $Lu_2MnCoO_6$ [15] and $Bi_2FeCrO_6$ [16].

Following Ref. [17], perovskites of oxide-type could be doped with titanium ions to modify their perovskite-type semi-conducting behavior as well as the other physical properties. Such perovskites, with semiconductor character, have drawn important interest as potential thermoelectric materials Ref. [18].

## 2. Computational details

$CaMnO_3$ crystallizes in the cubic perovskite structure $ABO_3$ with a lattice parameter of 3.73 Å, in the Pm3m space group. In cubic $CaMnO_3$, the Ca ion is surrounded by twelve oxygen ions. The smaller Mn ion is surrounded by six oxygen ions (see Figure 1). The atoms are positioned in the cubic perovskite of the calcium manganese oxide compound $CaMnO_3$ at Ca (0 0 0), Mn(1/2 1/2 1/2), and O (1/2 1/2 0). The equilibrium structural parameters and all properties considered in our calculation were computed using the Linearized Augmented Plane Wave (LAPW) method implemented in the Quantum Espresso code within the framework of density functional theory, using the GGA and GGA+ U approximations. The cut-off energy and k points are set at 50 Ry, 4x8x8 respectively to obtain a very good convergence of the total energy.

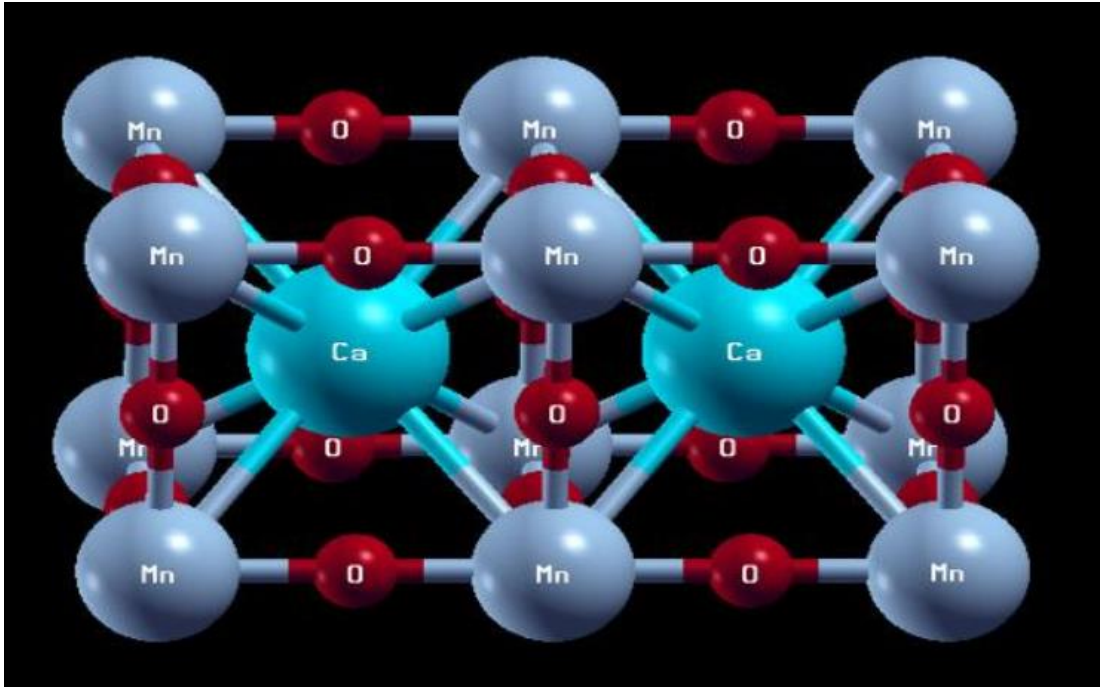

Figure 1: Representation of the perovskite structure CaMnO3.

### 3. Results and discussions

The compound CaMnO3 is known as a G-type antiferromagnetic semiconductor (AFM) with a Neel temperature TN=350K [41]. Our goal in this work is to study the ground state properties of CaMnO3. We have given the total energy as a function of the lattice parameter for both configurations ferromagnetic and anti-ferromagnetic (see Table. 1): The optimization cycle continues until convergence is achieved, which resulted in an equilibrium lattice parameter of a=3.70 Å. For more details, we have also given the total energy for the two configurations considered in our calculation for the cubic CaMnO3 with ferromagnetic (Ferro), and G-type anti-ferromagnetic (G-AFM) using GGA only (see Table. 1). The most stable magnetic state of the perovskite compound CaMnO3 corresponds to the phase with the lowest energy. We concluded that our compound CaMnO3 is G-type anti-ferromagnetic (G-AFM), which agrees very well with the theoretical work of Søndena et al. [3, 4], Pickett and Singh [19], Nicastro and Patterson [20].

| E(Ferromagnétique) | E(Antiferromagnétique) |
|---|---|
| −823.03026225 | −823.04067094 |

*Table 1: Total energy as a function of lattice parameter for the configurations ferromagnetic and anti-ferromagnetic.*

### 3.1. Magnetic moment

In this work, we examined especially the magnetic moment in all the atoms that formed our compound using the approximations GGA and GGA+U, respectively, for both configurations ferromagnetic and anti-ferromagnetic for the cubic CaMnO3 in Table 2. The introduction of the term U-Hubbard greatly influenced our results; This effect is more remarkable on the calculated magnetic moment of the manganese atom µ$_{Mn}$. Our calculated values obtained from the ferromagnetic configuration using GGA approximate, the value obtained by Søndena et al. ° [4], which is 2.47mB, and our calculated values obtained by GGA+U give 2.68mB, approximate the experimental work, which is equal to 2.6mB. The GGA+U is a more exact approximation than others utilized in our calculation.

### 3.2. Calculations of spins and magnetic moment theoretically:

The electronic structures of Calcium, manganese, and oxygen are:

Ca (z=20) : 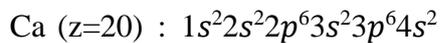 $1s^2 2s^2 2p^6 3s^2 3p^6 4s^2$

Mn (z=25) : 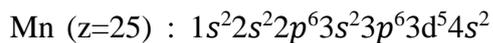 $1s^2 2s^2 2p^6 3s^2 3p^6 3d^5 4s^2$

O (z=8) : 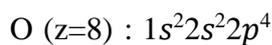 $1s^2 2s^2 2p^4$

❀ The valence electrons: O: $2s^2 2p^4$    Ca: $3p^6 4s^2$    Mn: $3d^5 4s^2$ 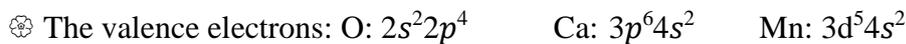

❀ The valence of O is $O^{2-}$:   $O^{2-}$: $2s^2 2p^6$ 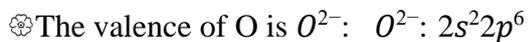 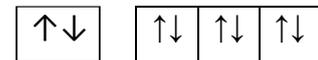

❀ The valence of Ca is $Ca^{2+}$:   $Ca^{2+}$: $3d^6 4s^0$ 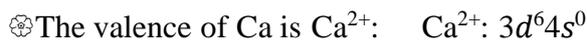 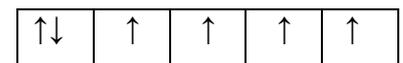

❀ The valence of Mn is $Mn^{4+}$:   $Mn^{4+}$: $3d^3 4s^0$ 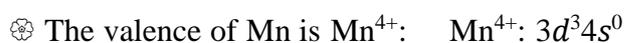 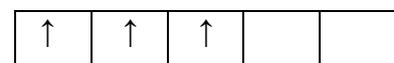

| | Spin moment | Orbital moment | Total moment | Magnetic moment |
|---|---|---|---|---|
| $Mn^{4+}$ | S = 3/2 | L = 3 | J = 3/2 | $\mu = 0.774 \mu_B$ |

**Table 2:** *Different types of moments (Spin, Orbital, Total, and Magnetic) for the $Mn^{4+}$.*

In fact, the total moment is $J = |L - S| = 3/2$; while the magnetic moment (effective) is:

$$\mu = g_J [J(J+1)]^{1/2} = 0.774 \ \mu_B \qquad (1)$$

where $g_J$ is the Landé factor. The different types of moments (Spin, Orbital, Total, and Magnetic) for the $Mn^{4+}$ are summarized in Table 2. In particular, at the saturation, the magnetic moment is given by:

$$\mu_{sat} = J \cdot g_J \cdot \mu_B = 0.6 \ \mu_B \qquad (2)$$

On the other hand, the Néel temperature, also known as Néel point, is the temperature above which an antiferromagnetic material becomes paramagnetic, meaning that the thermal energy is then sufficient to disrupt the microscopic magnetic order of the material. The energy level of the 3d electrons responsible for magnetism are slightly higher than that of the 2p and 4s conduction subshells.

## 4. Electronic properties of perovskite CaMnO3

### 4.1 Ferromagnetic phase:

We have shown the band Structure of the ferromagnetic cubic configuration in Fig. 2 and the G-type antiferromagnetic (G-AFM) cubic configuration in Fig. 3, which is more important and significant than the others. The total density of states (DOS) and partial density of states (PDOS) were calculated and plotted using the GGA and GGA+U approaches for the CaMnO3 compound in Figures 4, 5, 6, and 7. From the plot (Fig. 2), we can clearly notice that CaMnO3 exhibits metallic behavior for electrons with a spin-up due to the overlap of bands (the valence band and the conduction band), for electrons with a spin-down, CaMnO3 exhibits an insulating gap of 0.42 eV, with a maximum of the valence band at the point (M) and a minimum of the conduction band between M and Gama which means that the gap is indirect, which agrees with the result obtained by Søndena who gives the value of 0.46 eV [4] as a gap for the cubic structure of CaMnO3 compound. The band structures calculated for the Ferromagnetic configuration using GGA+U (**U=0.49 Ry**) show metallic behavior for the majority spin due to band overlap (the valence band and the conduction band). For the minority spin, CaMnO3 exhibits an insulating behavior with a direct band gap. Generally, the GGA+U (**U=0.49Ry**) gives a wider band gap than GGA. To better understand the band structure, it is interesting to determine the spectra of total and partial density of states in order to analyze and identify the type of hybridization, different interactions, and the states responsible for the bonding.

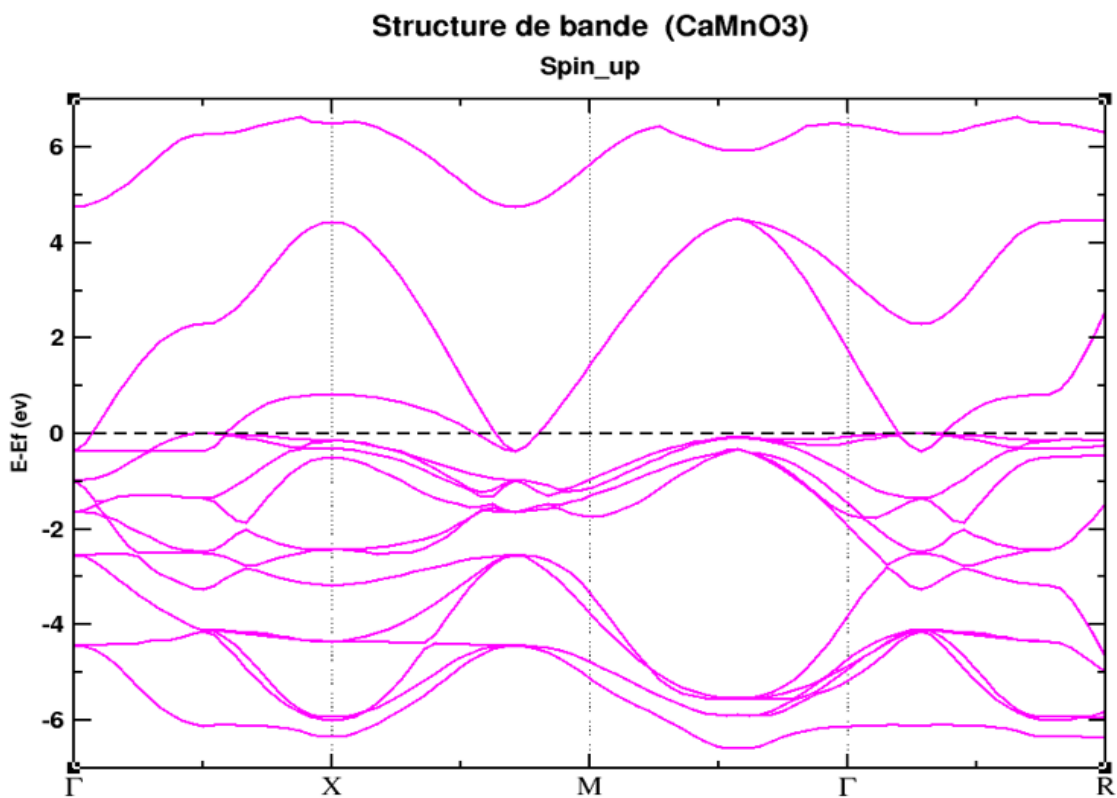

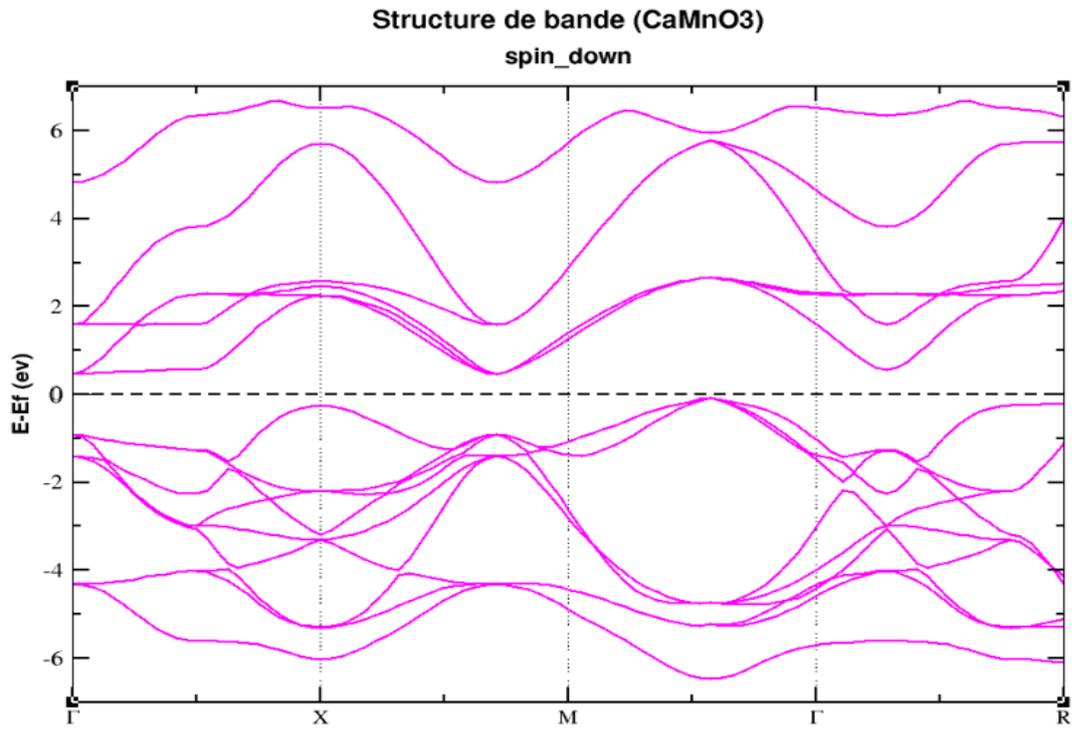

Figure 2: Band structures of CaMnO3 by GGA

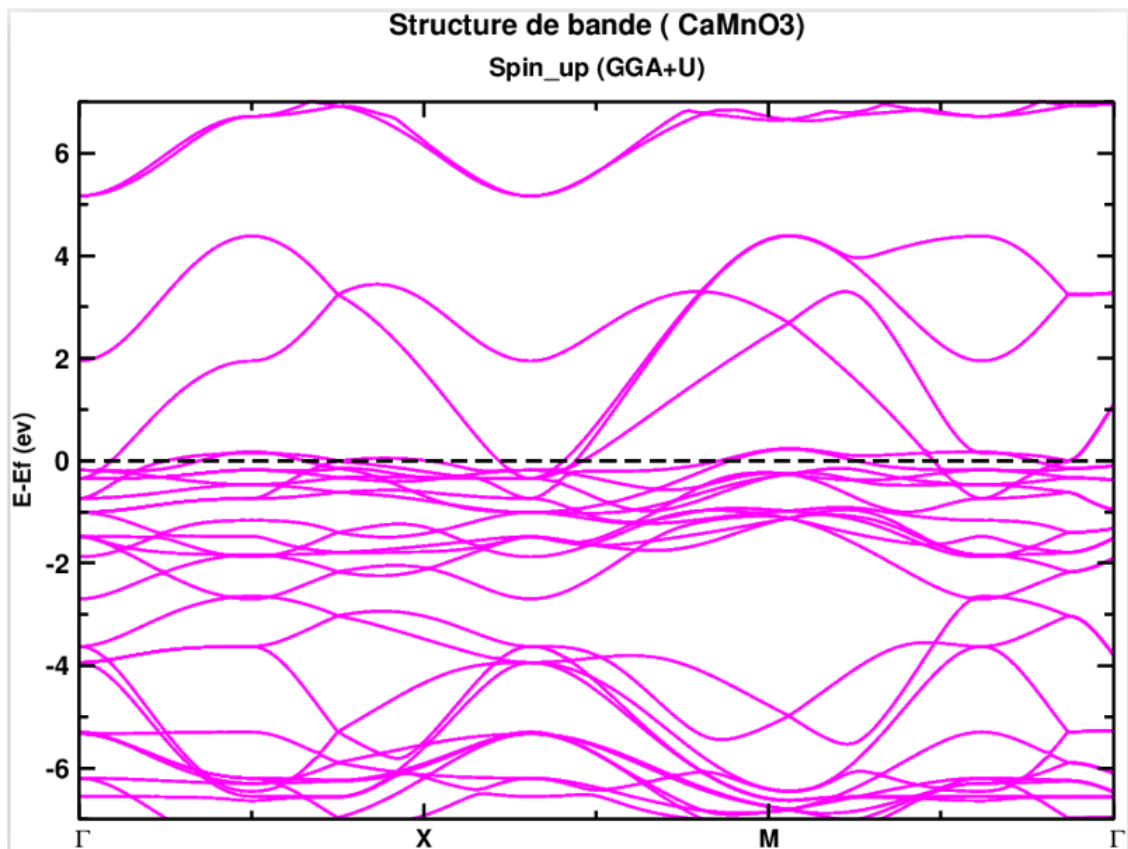

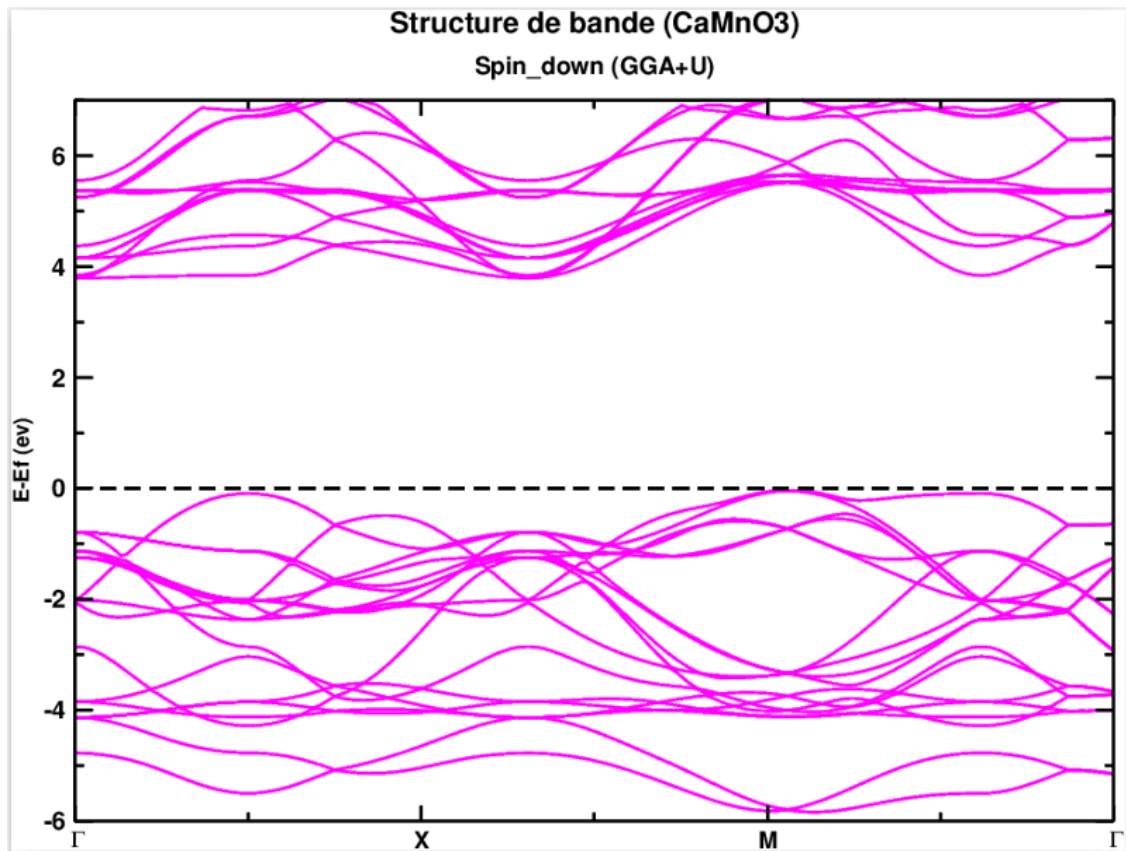

Figure 3: Band structures of CaMnO3 by GGA+U

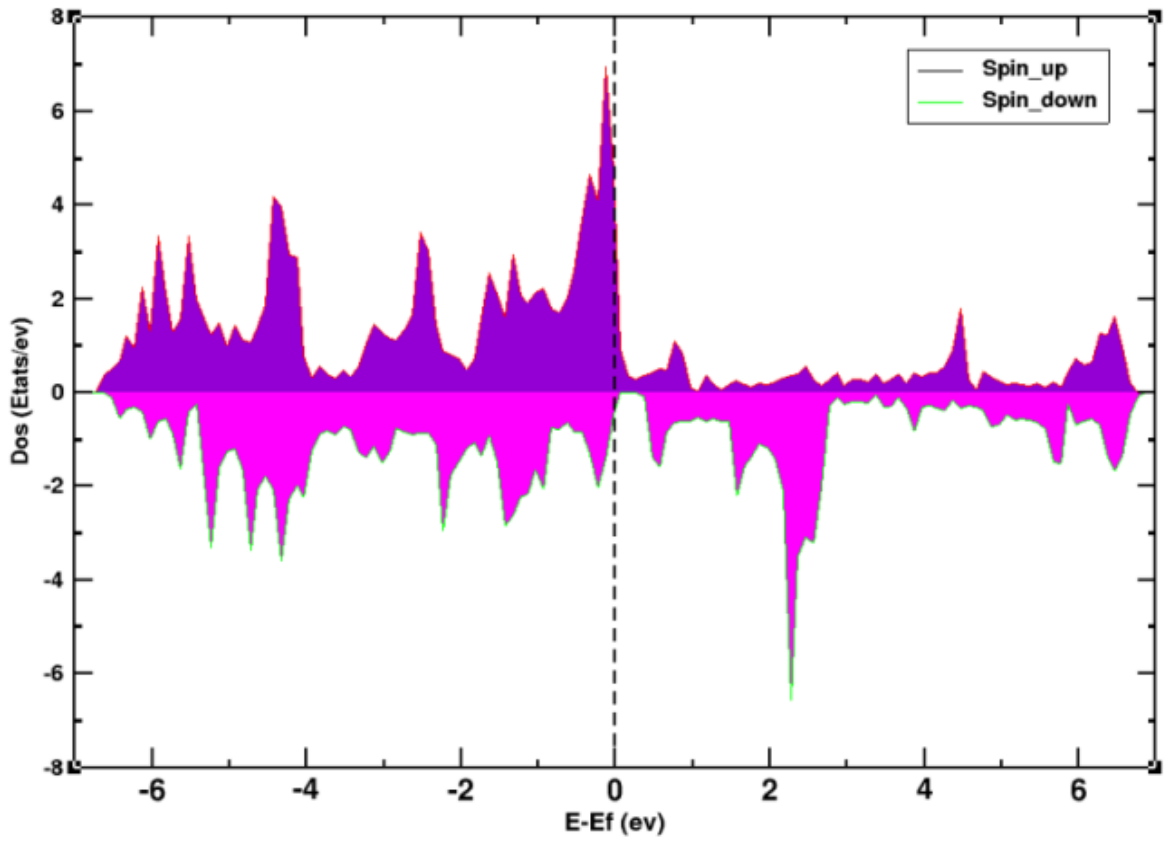

Figure 4: Total state density of CaMnO3 calculated by GGA.

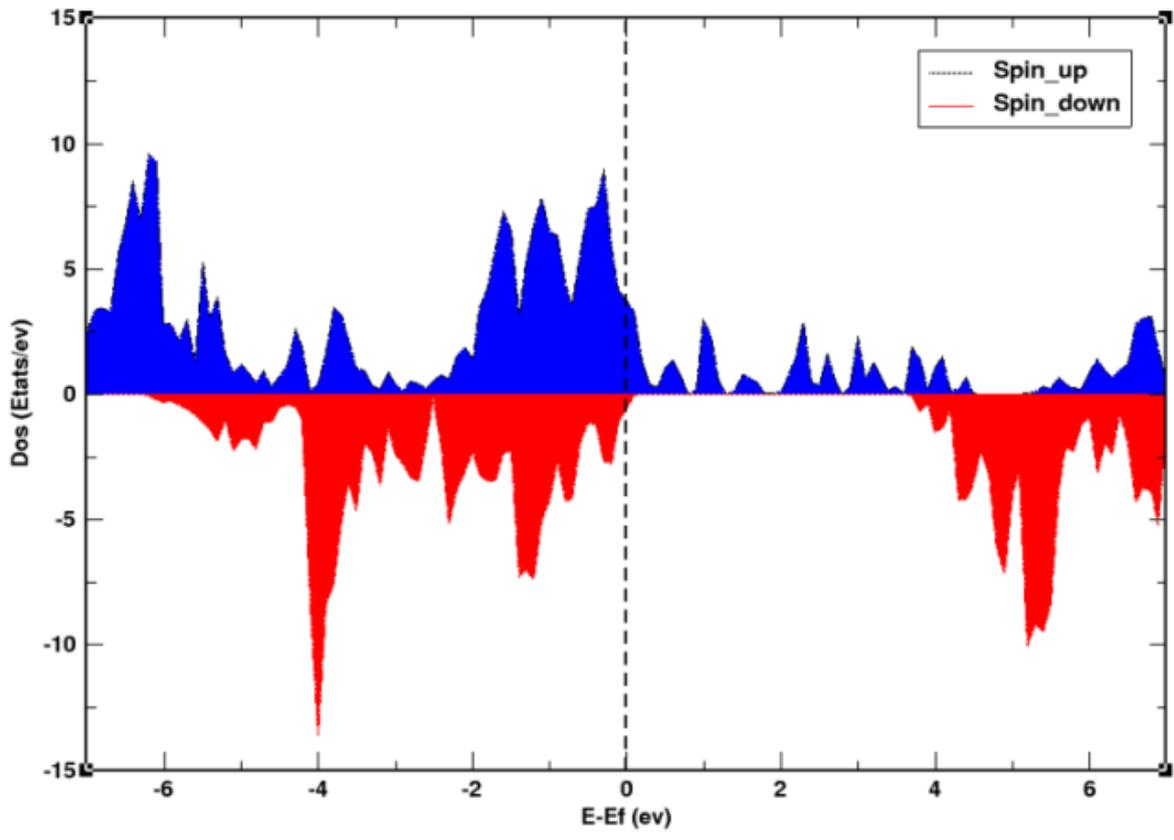

Figure 5: Total state density of CaMnO3 calculated by GGA+U (U=0.49Ry).

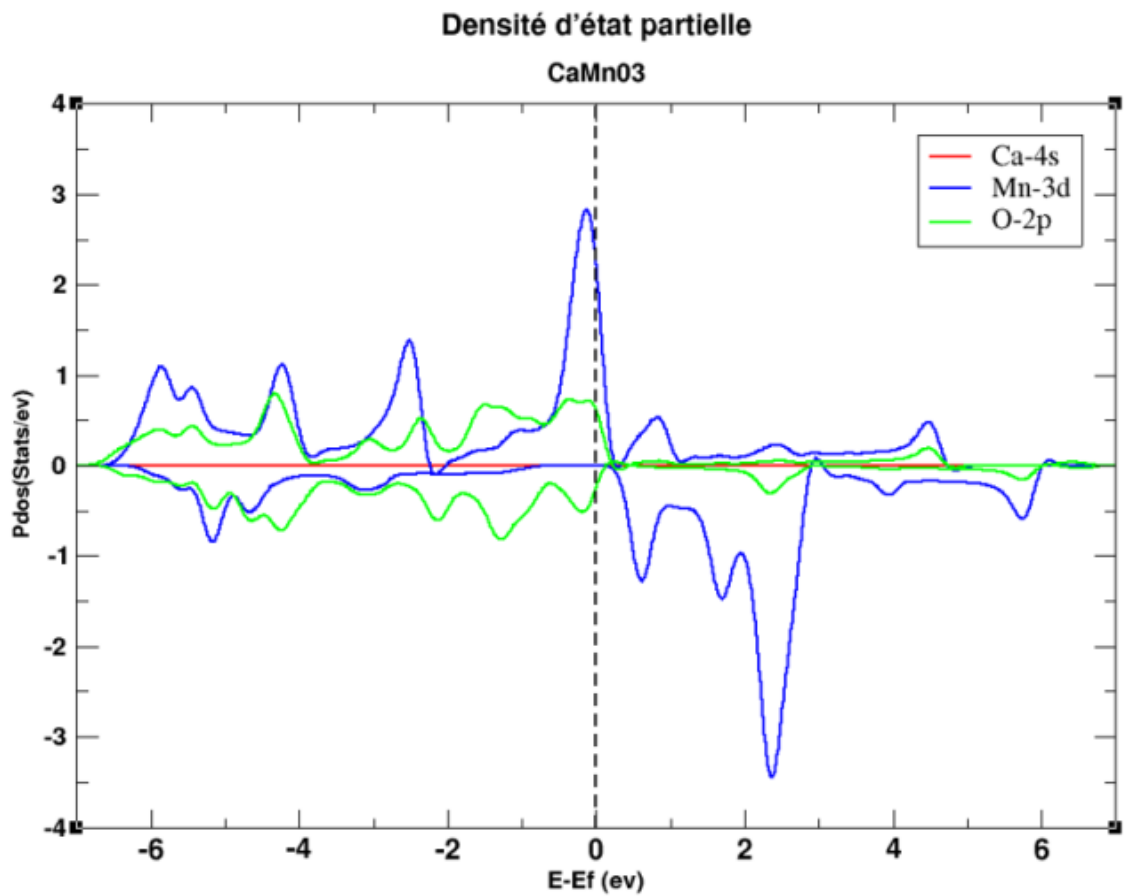

Figure 6: Partial state density of CaMnO3 calculated by GGA.

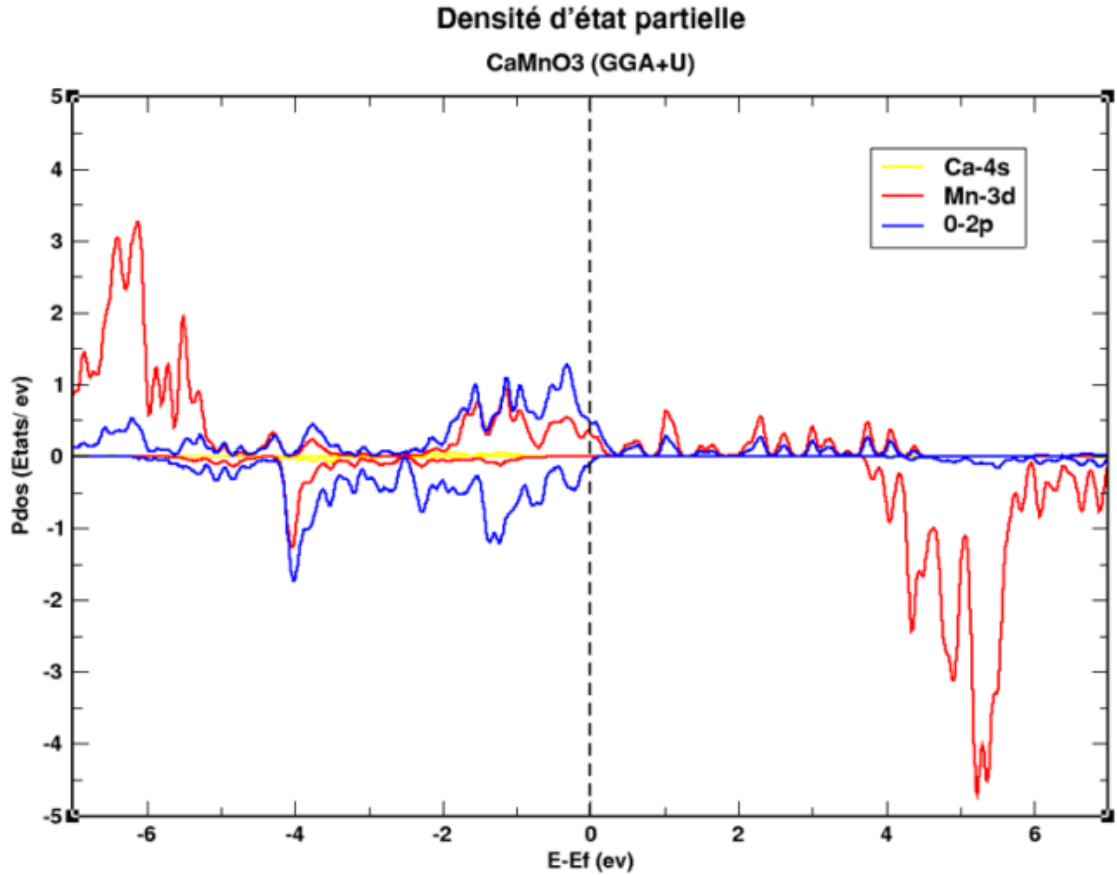

Figure 7: Partial state density of CaMnO3 calculated by GGA+U (U=0.49Ry).

The calculated density of states for CaMnO3 is depicted in Figures 4, 5, 6, and 7. The projected results cover the energy range from -7 eV to 7 eV, with the Fermi level (EF) set at an energy of 0 eV. The valence band (VB) spans from -7 eV to 0 eV, while the conduction band (CB) extends from 0 eV to 7 eV. We can see that the magnetic character is clearly visible in the density of states of our material. Indeed, the states of spins up and spin down are not symmetric. It can be observed that the electronic states around Fermi energy mainly come from Mn and O atoms and that the contribution of Ca atoms is very small. The resonance peaks in the electronic states of 'Mn' and 'O' atoms illustrate that there is a strong hybrid effect between the electrons of O-2p and Mn-3d.

**4.2 Antiferromagnetic phase:**

In the antiferromagnetic phase, the band structures relating to spins up are identical to those obtained for spins down. From the plot (Fig. 8 and 9), we can clearly observe that the Fermi level crosses an energy band, indicating the metallic behavior of CaMnO3 for both majority and minority spins using GGA and GGA+U.

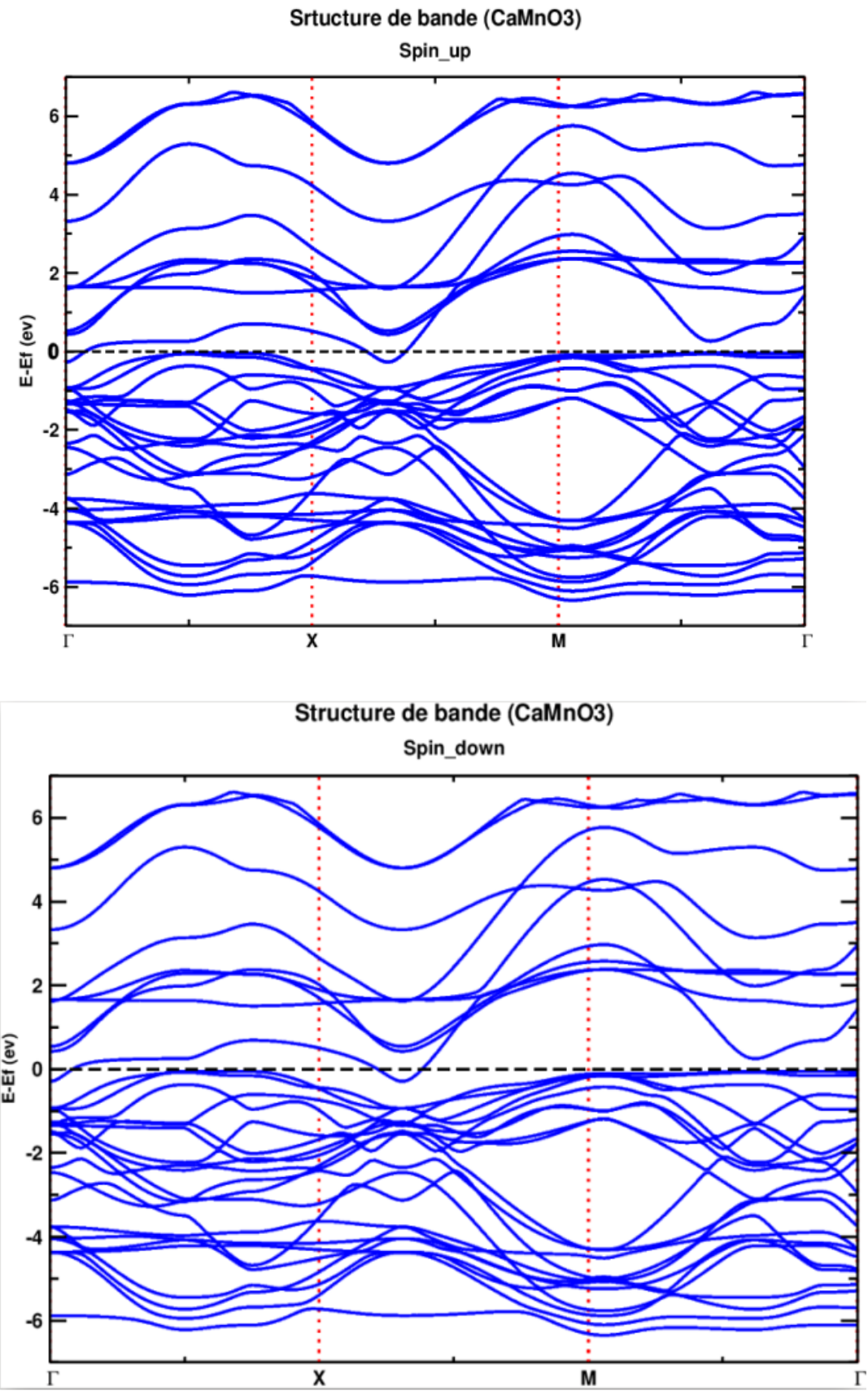

Figure 8: Band structures of the compound CaMnO3 by GGA.

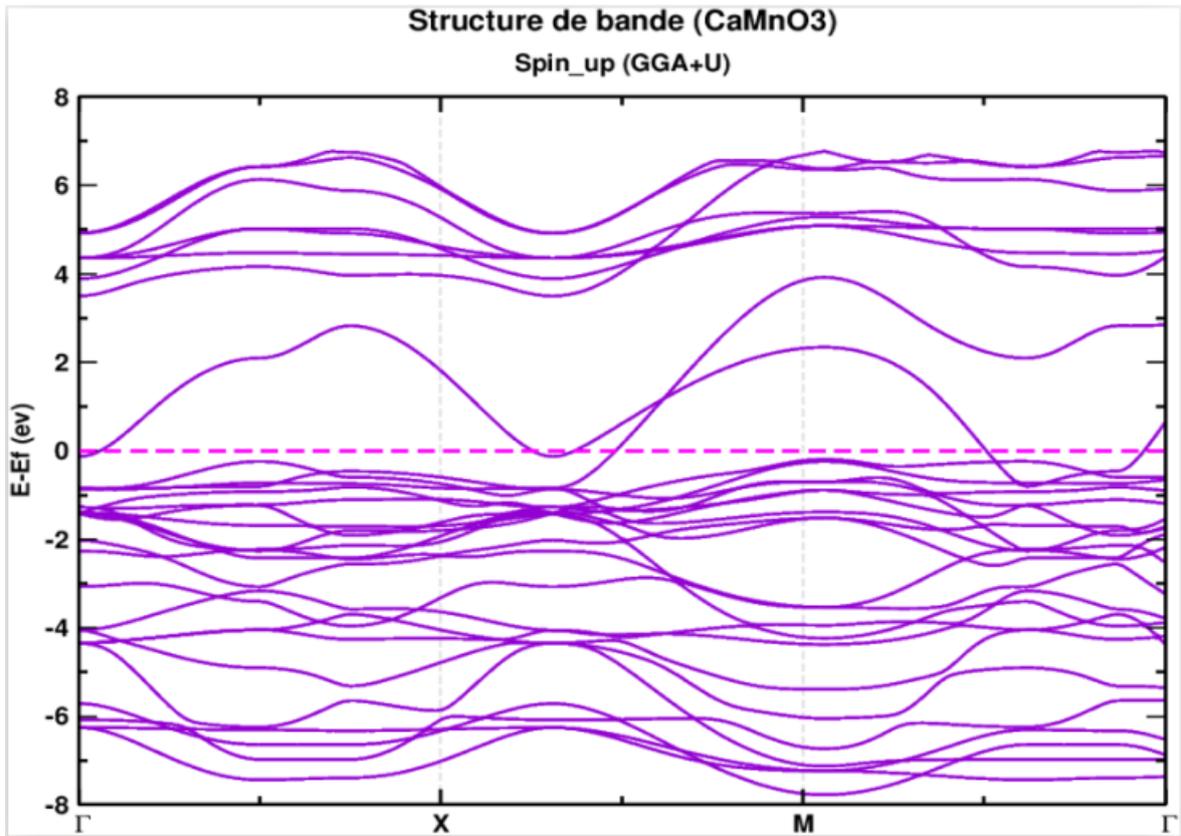

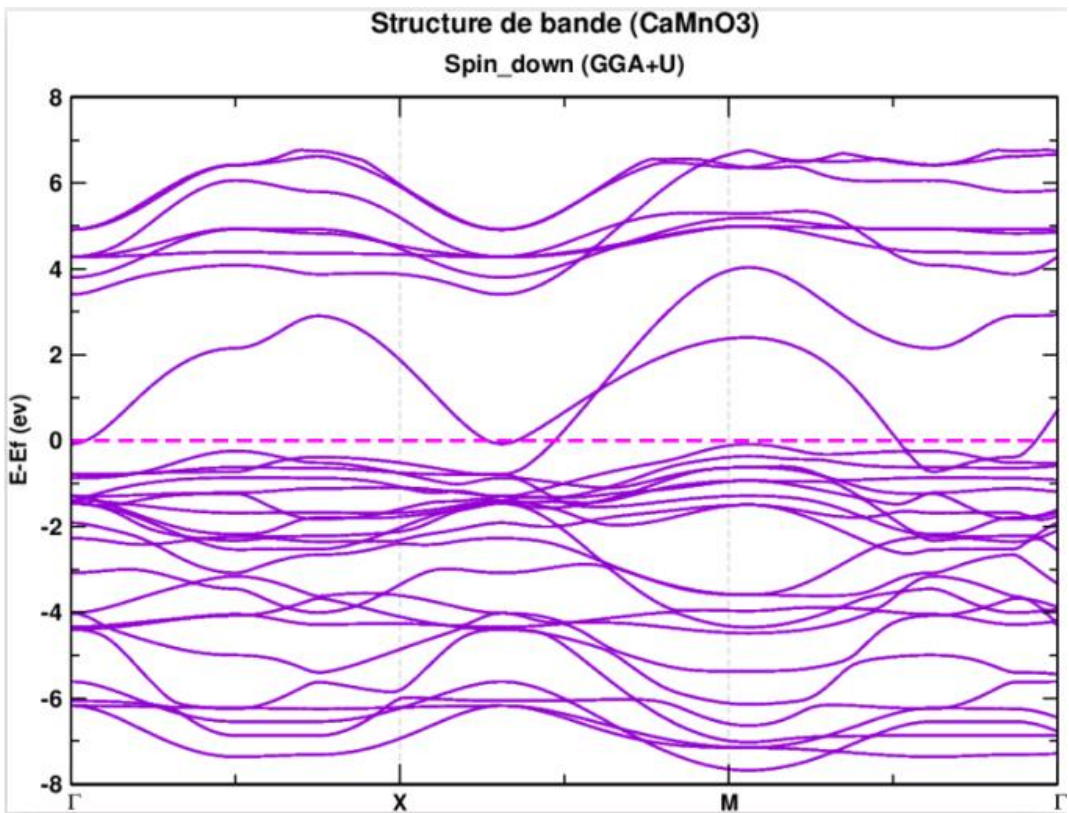

Figure 9: Band structures of the compound CaMnO3 by GGA+U (U=0.49Ry).

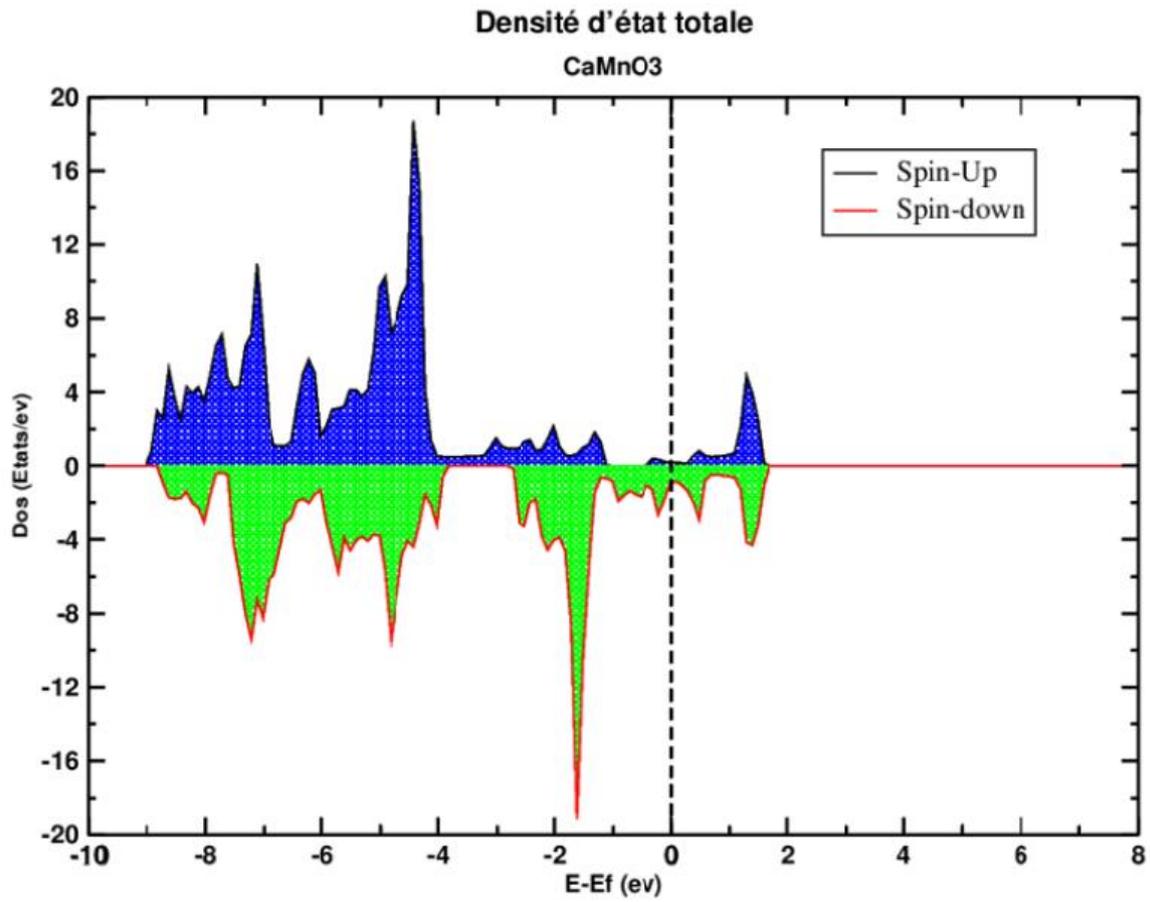

Figure 10: Total state density of CaMnO3 calculated by GGA.

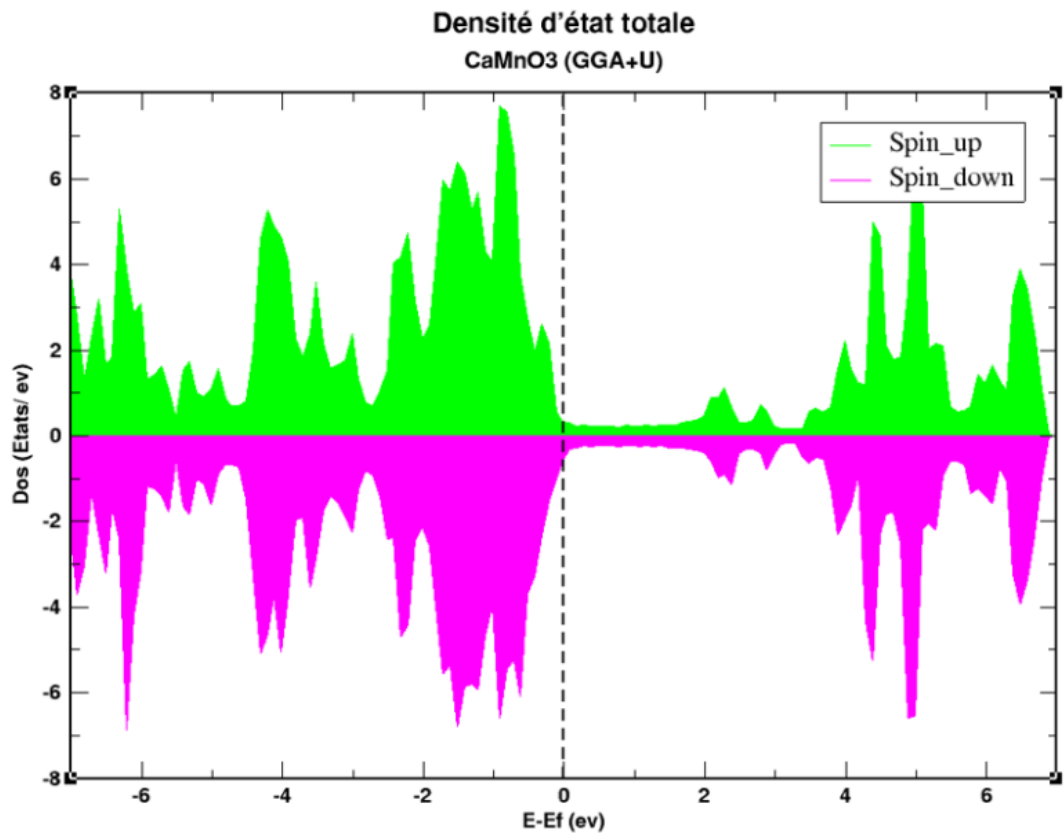

Figure 11: Total state density of CaMnO3 calculated by GGA+U (U=0.49Ry).

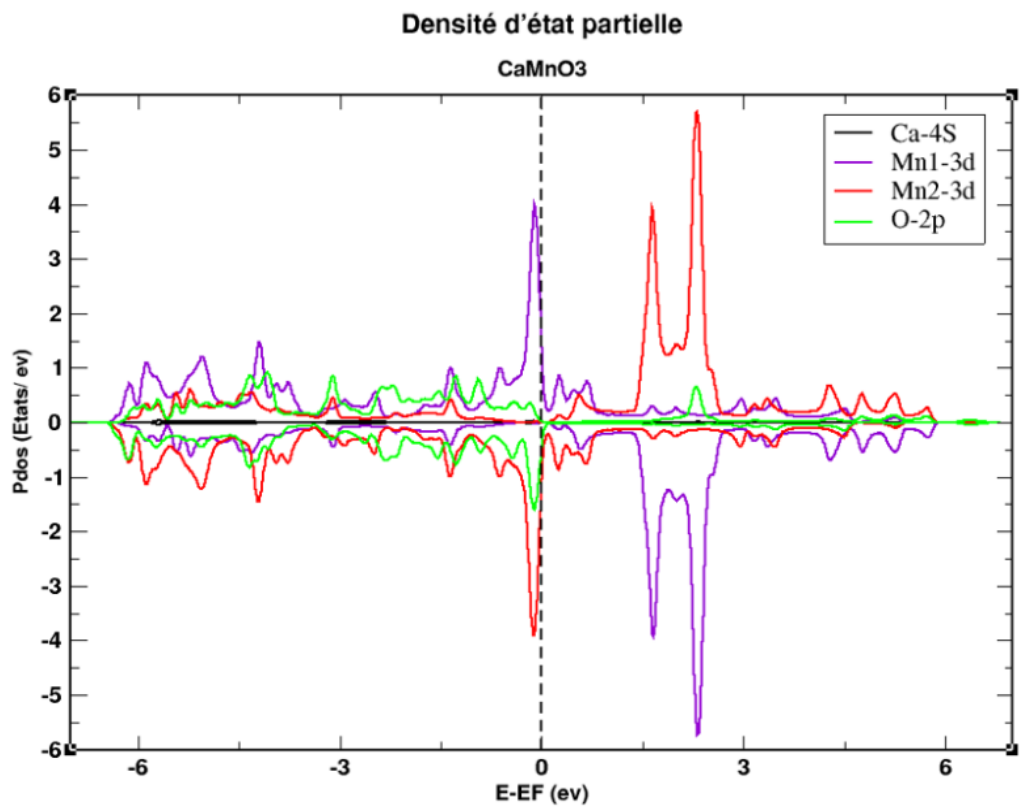

Figure 12: Partial state density of CaMnO3 calculated by GGA.

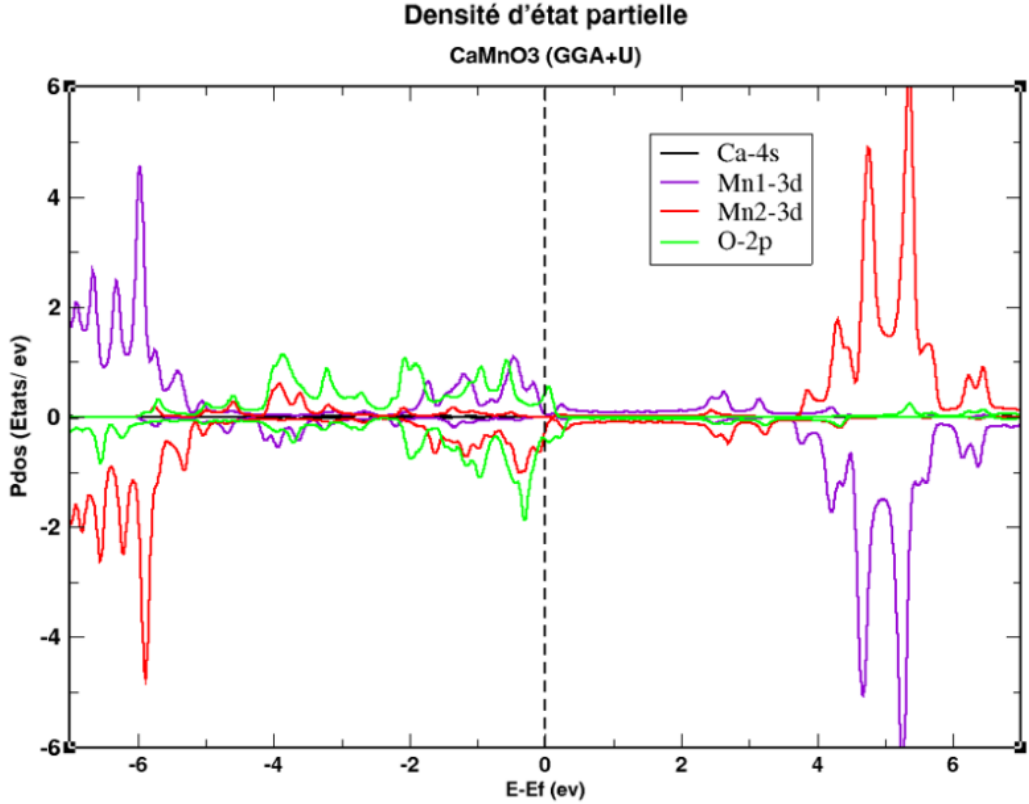

**Figure 13**: Partial state density of CaMnO3 calculated by GGA+U (U=0.49 Ry).

The calculated density of states of CaMnO3 is shown in Figures 10, 11,12, and 13. The projected results cover the energy range from -7 eV to 7 eV, Fermi level (EF) is taken at an energy of 0 eV. The valence band (VB) extends from -7 eV to 0 eV, and the conduction band (CB) extends from 0 eV to 7 eV. The density of states for the cubic G-AFM configuration of the compound CaMnO3 exhibits metallic behavior and a strong contribution of the Mn1 and Mn2 atoms. It also shows a small contribution from the p state of the oxygen atom. But it is more remarkable that GGA+U represents the density of states very well and shows very clearly the states of each atom, especially the s state of the calcium atom states, which are not very significant because the G-AFM DOS is dominated by the d-Mn states, which GGA+U much better presents than by GGA.

5. **Monte Carlo method**:

The Monte Carlo Approach is applied within the Ising model of the three-dimensional and the application of the Metropolis algorithm. The CaMnO3 compound's exchange parameters are categorized into different groups. Our focus is solely on the magnetic atoms, which are the Mn atoms. In this context, we are specifically looking at J1, which represents the exchange parameter between Mn atoms on nearest neighbor sites. The Hamiltonian model of the YTiO3 compound was given as follows:

$$H = -J_1 \sum_{<i,j>} S_i^{Mni} S_j^{Mni} \qquad (1)$$

The first nearest neighbors in the CaMnO3 compound consist of 6 atoms. J1 is the coupling interaction between Mn atoms (first nearest neighbors). The dominant exchange interactions among the first nearest neighbors, J1, were calculated using the conventional total energy technique, resulting from: J1= 5.242592593 meV.

6. Results achieved by the Monte-Carlo program :

To present the Monte Carlo results, we worked with a cube lattice consisting of a single magnetic atom located at the vertices of a cube.

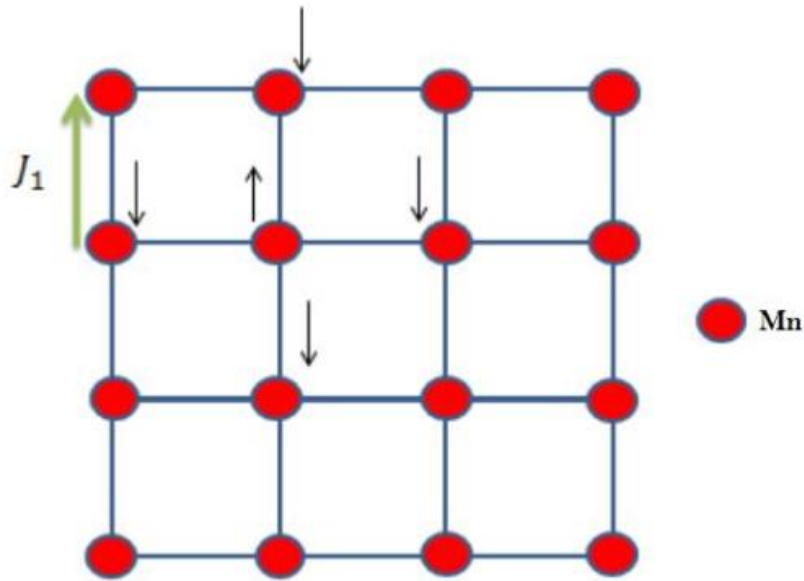

Fig. 14: The filling of the atoms in the lattice of the CaMnO3 structure.

**6.1 Magnetizations:**

The figure below represents the magnetization of the system as a function of temperature for different sizes. The curves clearly indicate that it is a second-order phase transition from the Antiferromagnetic phase to the Paramagnetic phase. The variation in size shows a convergence of the results for sufficiently large sizes. However, in the literature, the antiferromagnetic to paramagnetic transition occurs at TN = 350 K [21], and the difference between this value and that one obtained in the calculation can be explained by the coupling value calculated at the DFT level.

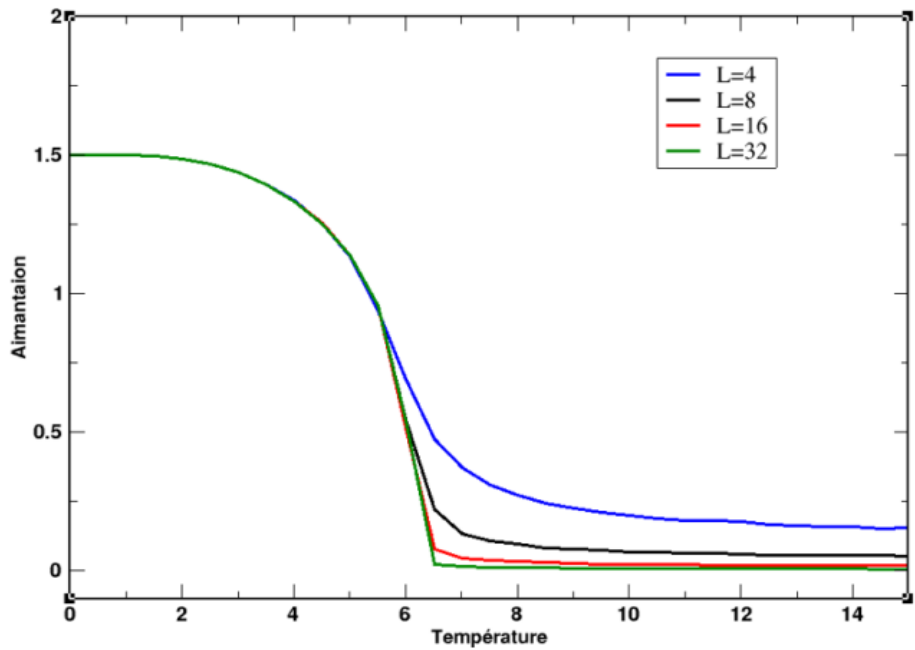

**Fig. 15**: Evolution of the magnetization as a function of the temperature.

### 6.2 Magnetic susceptibility:

The measurement of susceptibility is based on the fluctuations in the magnetization of the system. The figure below illustrates the variation of susceptibility as a function of temperature. At low temperatures, the susceptibility is null because the magnetization is constant. We observe that susceptibility is represented by peaks since we are dealing with finite systems, not at the thermodynamic limit. The transition temperature is compatible with that found by magnetization. The susceptibility diverges in the vicinity of the Néel temperature.

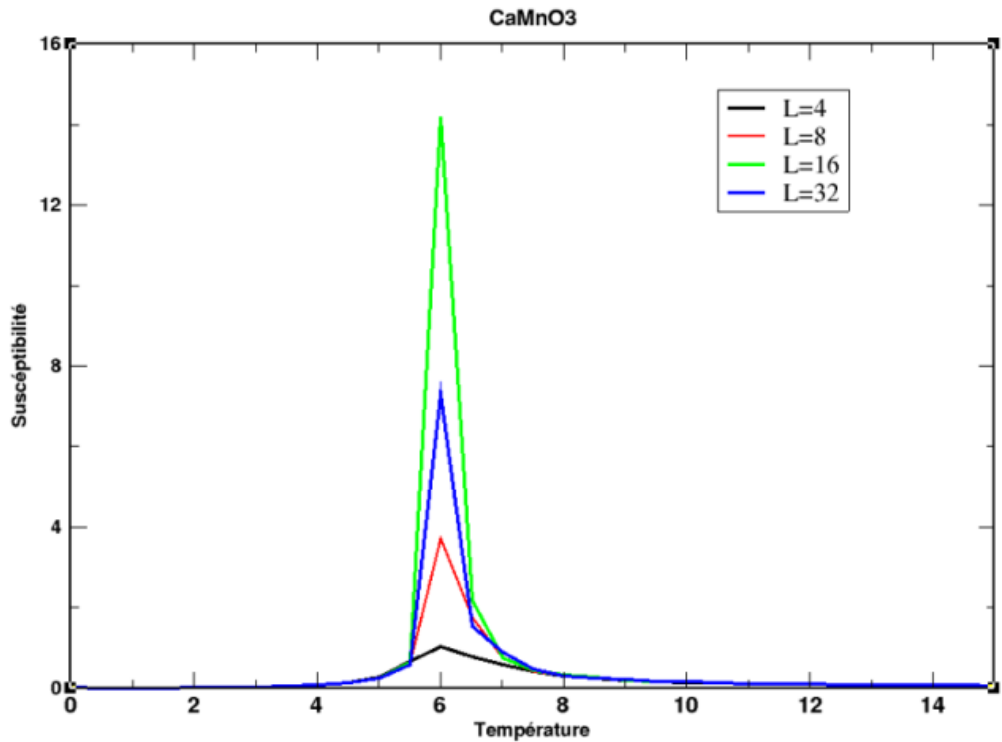

**Figure 16:** Evolution of susceptibility as a function of temperature for L=4, 8, 16 and 32.

### 6.3 Specific heat:

The specific heat is a parameter that depends strongly on the temperature and energy of the system. The variation of this parameter is represented in Fig. 17, as a function of temperature for different system sizes. The specific heat increases until it reaches a maximum value corresponding to the Néel temperature, after which it decreases at high temperatures. This parameter diverges near the Néel temperature.

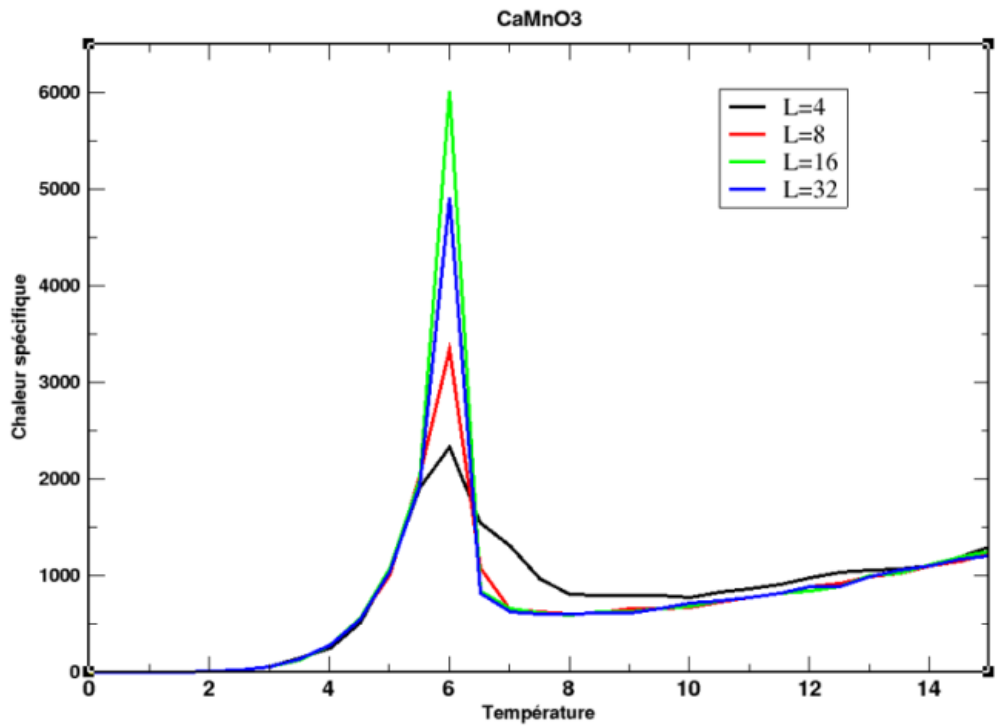

Fig. 17 Shows the evolution of the specific heat as a function of the temperature for L=4, 8, 16, and 32.

### 6.4 Binder's Cumulant

The Binder cumulant (BC) is a more numerically efficient method for determining the Néel temperature. From the figure below we can confirm that when we change the size of the system all the points converge to the same point, which is the transition point.

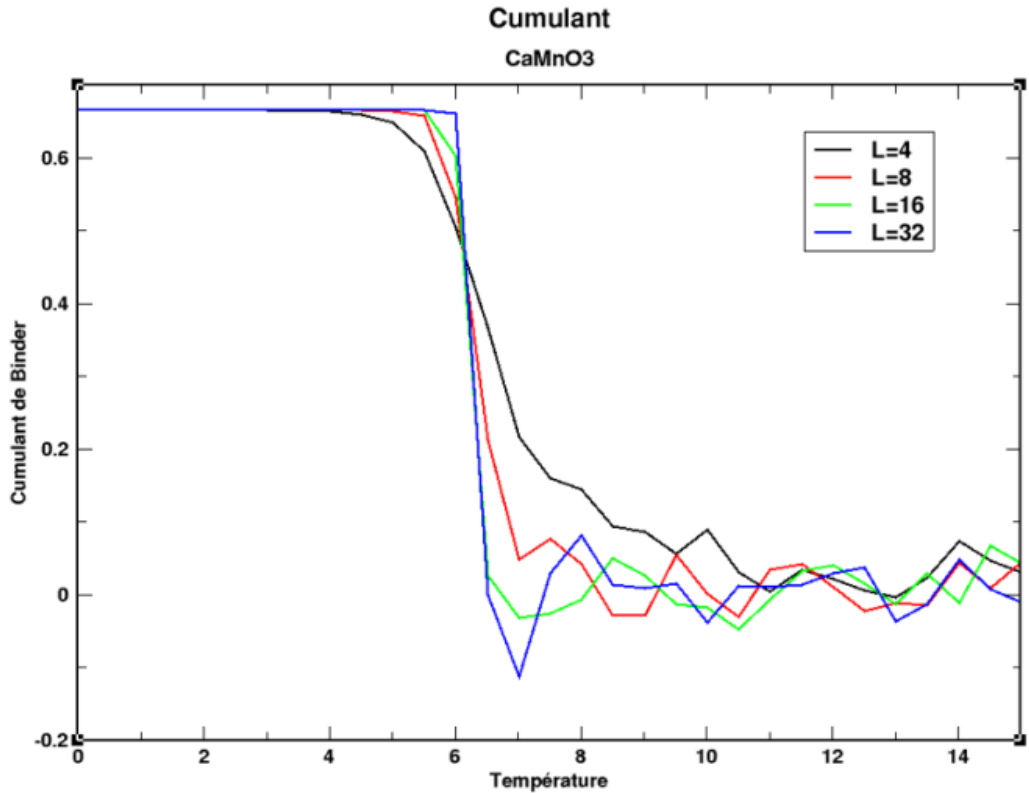

**Fig. 18** Evolution of the Binder Cumulant (BC) as a function of temperature.

At high temperatures, the fluctuations observed are due to the Monte Carlo simulations. The numerical value of the exchange coupling interaction is J1= 0.005242592593 eV. The corresponding Néel temperature value is $TN = 6\frac{J}{K_B} = 364.7$ K. This value is close to the experimental one TN=350 K provided in Ref. [8].

7. **Conclusion**

we used an ab initio density functional calculation using the approaches (GGA, GGA+U) to evaluate the electronic and magnetic properties of the ideal cubic perovskite CaMnO3. These properties are constantly being exploited in different areas of our daily lives. It has been inferred that CaMnO3 is stable in the G-AFM phase. From the total and partial DOS, we found a strong contribution of the d-Mn states. We have implemented the exchange and correlation potentials to compare the two approximations GGA and GGA+U, with the last one being meant with onsite Coulomb interaction U and the site exchange interaction J to treat the localized d electron states in the manganese atom. This choice results in an agreement between the calculated magnetic moment and the theoretical and experimental data. The GGA+U gives more accurate results than others give and shows the bonding between atoms with efficiency. In summary,

Quantum Espresso is a powerful tool for performing ab initio calculations and determining various physical quantities for a wide variety of materials and systems. The simulation using Monte Carlo code based on the Metropolis algorithm helped us to simulate the physical (compound) system with finite T on a machine and to observe the evolution of the three main quantities: magnetization, susceptibility, and specific heat of the material.